\documentclass{optica-article}
\usepackage{ulem,xpatch}
\journal{opticajournal} 

\articletype{Research Article}
\usepackage{soul}
\usepackage{lineno}

\begin{document}

\title{SiN integrated photonic components in the Visible to Near-Infrared spectral region}
\author{Matteo Sanna,\authormark{1,*} Alessio Baldazzi,\authormark{1} Gioele Piccoli,\authormark{1,2} Stefano Azzini,\authormark{1} Mher Ghulinyan,\authormark{2} and Lorenzo Pavesi\authormark{1}}

\address{\authormark{1}Department of Physics, University of Trento, Via Sommarive 14, 38123, Trento, Italy\\
\authormark{2}Centre for Sensors and Devices, Fondazione Bruno Kessler, via Sommarive 18, 38123, Trento, Italy}

\email{\authormark{*}matteo.sanna@unitn.it} 


\begin{abstract*} 
Integrated photonics has emerged as one of the most promising platforms for quantum applications. The performances of quantum photonic integrated circuits (QPIC) necessitate a demanding optimization to achieve enhanced properties and tailored characteristics with more stringent requirements with respect to their classical counterparts. In this study, we report on the simulation, fabrication, and characterization of a series of fundamental components for photons manipulation in QPIC based on silicon nitride. These include crossing waveguides, multimode-interferometer-based integrated beam splitters (MMIs), asymmetric integrated Mach-Zehnder interferometers (MZIs) based on MMIs, and micro-ring resonators. Our investigation revolves primarily around the Visible to Near-Infrared spectral region, as these devices are meticulously designed and tailored for optimal operation within this wavelength range. By advancing the development of these elementary building blocks, we aim to pave the way for significant improvements in QPIC in a spectral region only little explored so far.

\end{abstract*}

\section{Introduction}
\label{sec:intro}

Advancements in quantum technologies over the past decade have the potential to revolutionize sensing \cite{degen2017quantum}, communication \cite{gisin2007quantum} and computation \cite{DiVinc, nielsen2001quantum,arute2019quantum}. Quantum applications do not have an \textit{a-priori} preferred material platform, since each of these has its strengths and weaknesses. This is especially true for integrated quantum photonics, whose architecture is based on the manipulation of photon states through integrated optical circuits. 
For instance, measurement-based quantum computing \cite{raussendorf2001a,briege2009measurement-based,bombin2021modular} implemented on silicon photonics \cite{rudolph2017optimistic} has gained much attention because of the impressive advancements in the integration of the elementary building blocks in a material platform which is compatible with electronic circuits. Such an interest is generated by the idea of a fault-tolerant quantum computer based on linear optical elements \cite{wang18,reimer19,zhong20,viglia2021error-protected}. Another example is that of Boson Samplers integrated in the direct-laser written glass platform, which allows demonstrating the ability of photonic structures to perform computational tasks beyond classical capabilities \cite{boson_sampling, boson_sampling_review}.

In the context of QPIC, silicon nitride (SiN) is gaining much attraction due to its exceptional optical properties. This material leverages CMOS technology for its fabrication, achieving the remarkable feature of producing low propagation loss QPICs from inherently compact dimensions. Notably, due to the large optical bandgap of $\sim$5 eV at the visible to near-infrared spectral region, SiN does not suffer from two-photon absorption (TPA), a phenomenon that constrains the efficiency of nonlinear applications in silicon waveguides. SiN possesses a nonlinear coefficient $n_2$ of 2.4 x $10^{-19}$ $m^2$/W \cite{ikeda2008thermal}, a value one order of magnitude lower than that of silicon, yet positioning SiN as a formidable contender for quantum-oriented applications \cite{cernansky2018complementary}. Another advantage of SiN is its spectral coverage, spanning from 0.35 $\mu$m to 7 $\mu$m with negligible material absorption (or up to 3.0 $\mu$m when embedded in silica)\cite{8472140,piccoli2022silicon}. This distinctive feature positions SiN as an optimal choice for QPICs within the visible spectral range, where silicon is not transparent\cite{cernansky2018complementary,leone2023generation,bres2023supercontinuum}, enabling the use of silicon for the direct integration of single photon visible detectors \cite{bernard2021top}. Moreover, this broad operation window makes SiN also compatible with single-photon sources based on a wide variety of Quantum Dots semiconductor compounds \cite{shan2014single}, for example around $800-815$ nm \cite{PhysRevB.72.195332,malko2006optimization} or around $735-790$ nm \cite{kimura2005single,kiravittaya2006ordered}.

This paper presents the performances of state-of-the-art SiN-based photonic integrated circuits (PICs) and discusses their potential use in both classical and quantum applications. Specifically, it will discuss basic photonic components for manipulating classical and quantum states of light such as crossing waveguides, beam splitters based on multi-mode interferometers (MMIs), photons filters and routers via asymmetric Mach-Zehnder interferometers (aMZI) and micro-ring resonators in a spectral region between 650 and 850 nm. To the best of our knowledge, all these components together in such spectral region have not been investigated so far. 
The paper is organized as follows. Section 2 provides a comprehensive account of the design as well as fabrication methods employed for the realization of SiN-based components. In Section 3, both theoretical and experimental aspects of waveguide crossings, MMIs, aMZIs, and micro-ring resonators are presented. \\

\section{Waveguides}
\label{sec:fab}

\begin{figure}[htbp]
\centering
\includegraphics[scale=0.39]{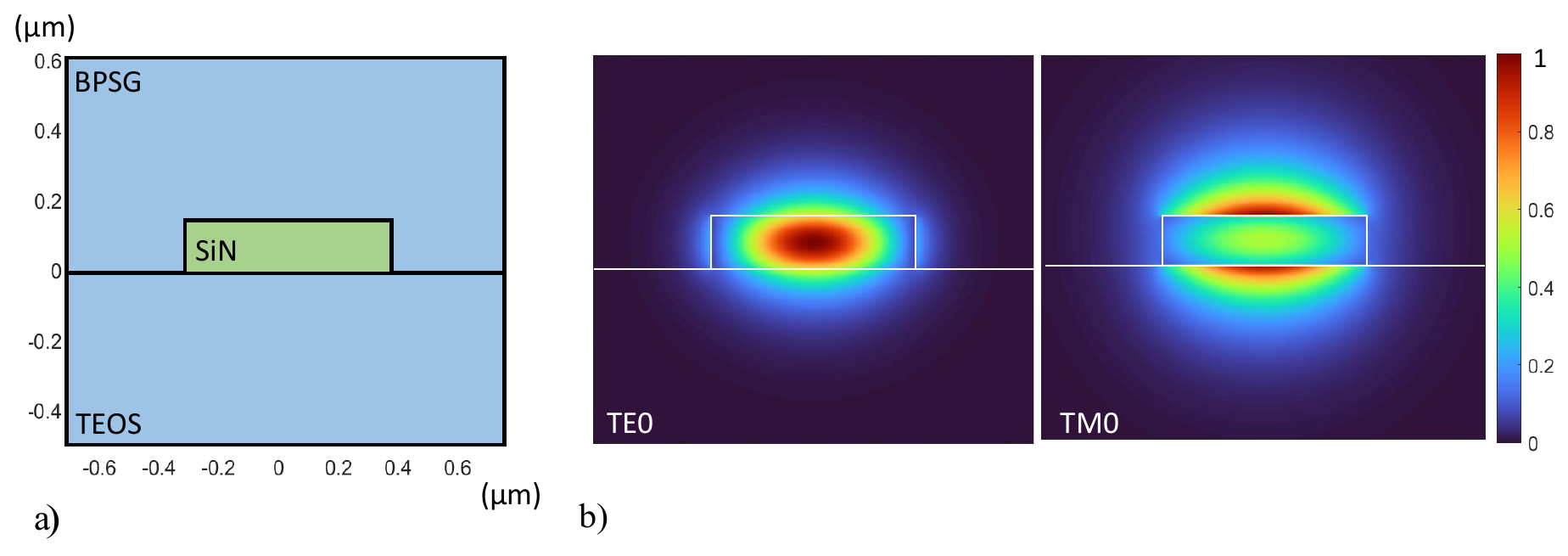}
\caption{a) Cross-section of the SiN core waveguide ($n_{SiN} = 1.991$ at 750 nm). The cladding consists of two distinct materials, namely borophosphosilicate glass (BPSG) on top of the waveguide ($n_{BPSG} = 1.459$ at 750 nm) and tetraethoxysilane (TEOS) based silicon oxide as substrate ($n_{TEOS} = 1.441$ at 750 nm). b) Simulated electric-field intensity profiles at 750 nm for the fundamental transverse-electric (TE) and transverse-magnetic (TM) modes.}
\label{mode}
\end{figure}

PICs were realized in stoichiometric silicon nitride (SiN) deposited via low-pressure chemical vapor deposition (LPCVD) on $150$~mm diameter silicon wafer in the FBK's cleanroom facilities. First, a bottom SiO$_2$ cladding of $1.7~\mu$m was grown using tetraethoxysilane (TEOS) gas precursor at $710$~\textdegree C chamber temperature, directly followed by the $140$~nm-thick SiN film deposition at $770$~\textdegree C. Next, the photonic circuits were patterned on standard photoresist using i-line stepper lithography, and directly defined in the SiN using an inductively-coupled plasma reactive ion etching. Finally, the realized SiN circuits were covered with a top, $1.6~\mu$m thick, borophosphosilicate glass (BPSG) cladding at $640$~\textdegree C, and the wafer was diced into single chips. The SiN channel waveguides have refractive index $n_{SiN} = 1.991$  at 750 nm and a $140$~nm $\times$ $650$~nm cross-section (Fig.~\ref{mode}(a)). The dimensions of the SiN waveguides were chosen to achieve single-mode conditions at 750 nm for both the transverse electric (TE) and transverse magnetic (TM) polarizations \cite{8472140}, with a mode field distribution as shown in Fig.~\ref{mode}(b).

\begin{figure}[htbp]
\centering
\includegraphics[scale=0.45]{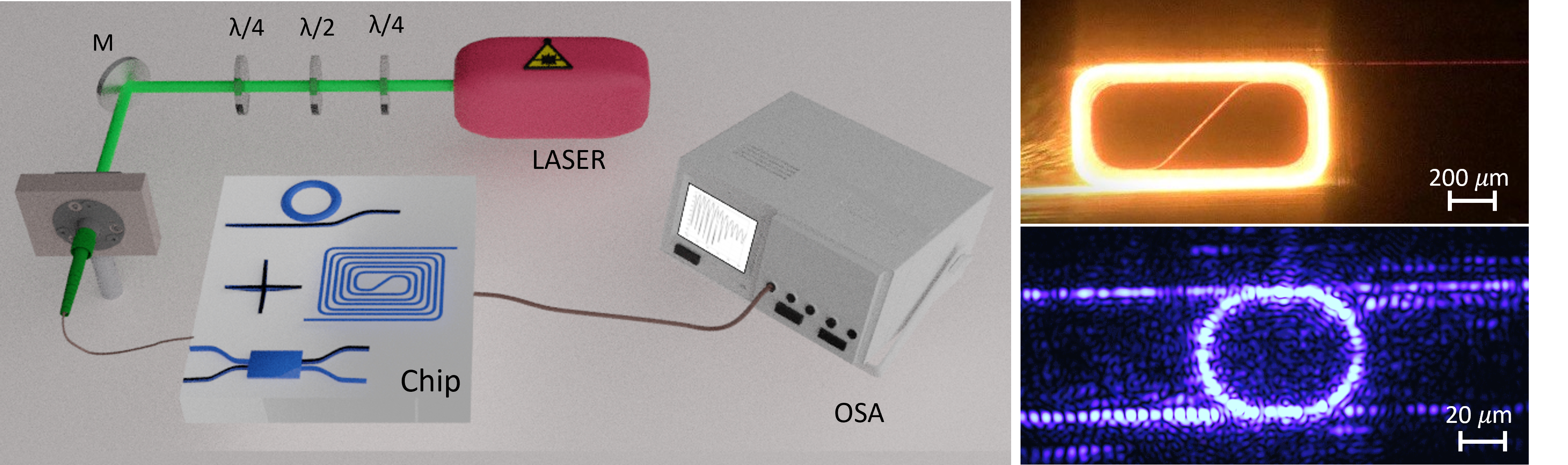}
\caption{On the left, a sketch of the experimental setup used for the characterization of the SiN integrated components. A supercontinuum laser is used as light source and is coupled with the PIC through a single-mode tapered fiber with a 1.25 $\mu$m spot size. The desired polarization state is set by utilizing free-space wave-plates $\lambda/4$ and $\lambda/2$. At the output of the PIC, the signal is collected by another single-mode tapered fiber and measured with an optical spectrum analyzer (OSA) allowing for a high-resolution spectral analysis of the optical signals. On the right, images taken by a camera of the light scattered from the top surface of the PIC by a waveguide spiral and a micro-ring during measurement.} 
\label{setup}
\end{figure}

Linear characterization measurements were conducted using the experimental setup sketched in Fig.~\ref{setup}. The setup incorporated a supercontinuum laser (FYLA SCT500) as the light source, which enabled the measurement of a wide-ranging spectral response covering wavelengths from 650 to 850 nm. The input polarization is determined by employing achromatic half-wave and quarter-wave plates. The output spectra were acquired using an Optical Spectrum Analyzer (OSA, Yokogawa-AQ6373B), allowing for high-resolution (100 pm) characterization of the optical signals emitted by the supercontinuum laser.

\begin{figure}[ht!]
    \begin{minipage}{0.49\textwidth}
    \centering
    {\small a)}\includegraphics[scale=0.31]{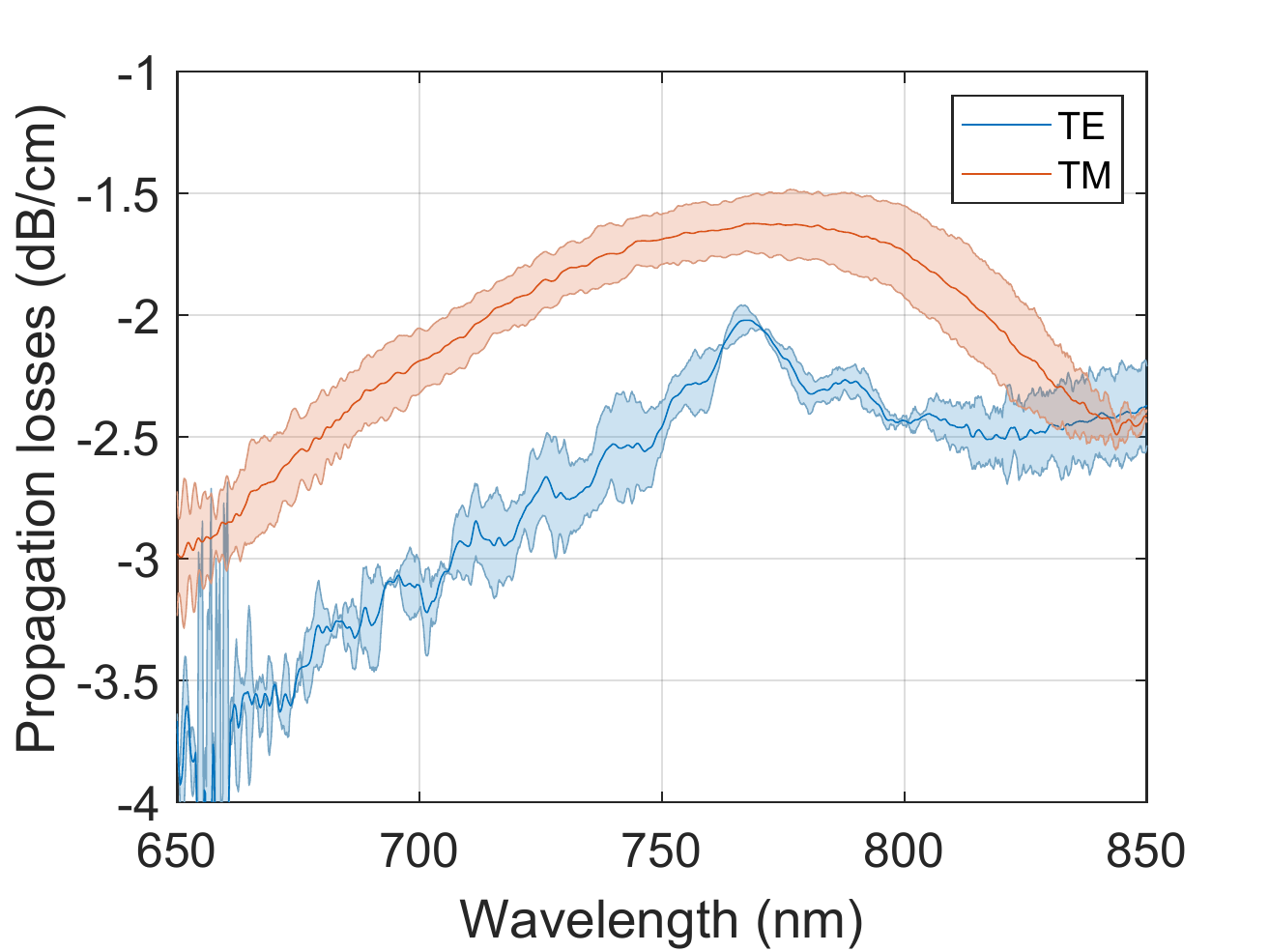}
    \end{minipage}
    \begin{minipage}{0.49\textwidth}
   \centering
    {\small b)}\includegraphics[scale=0.31]{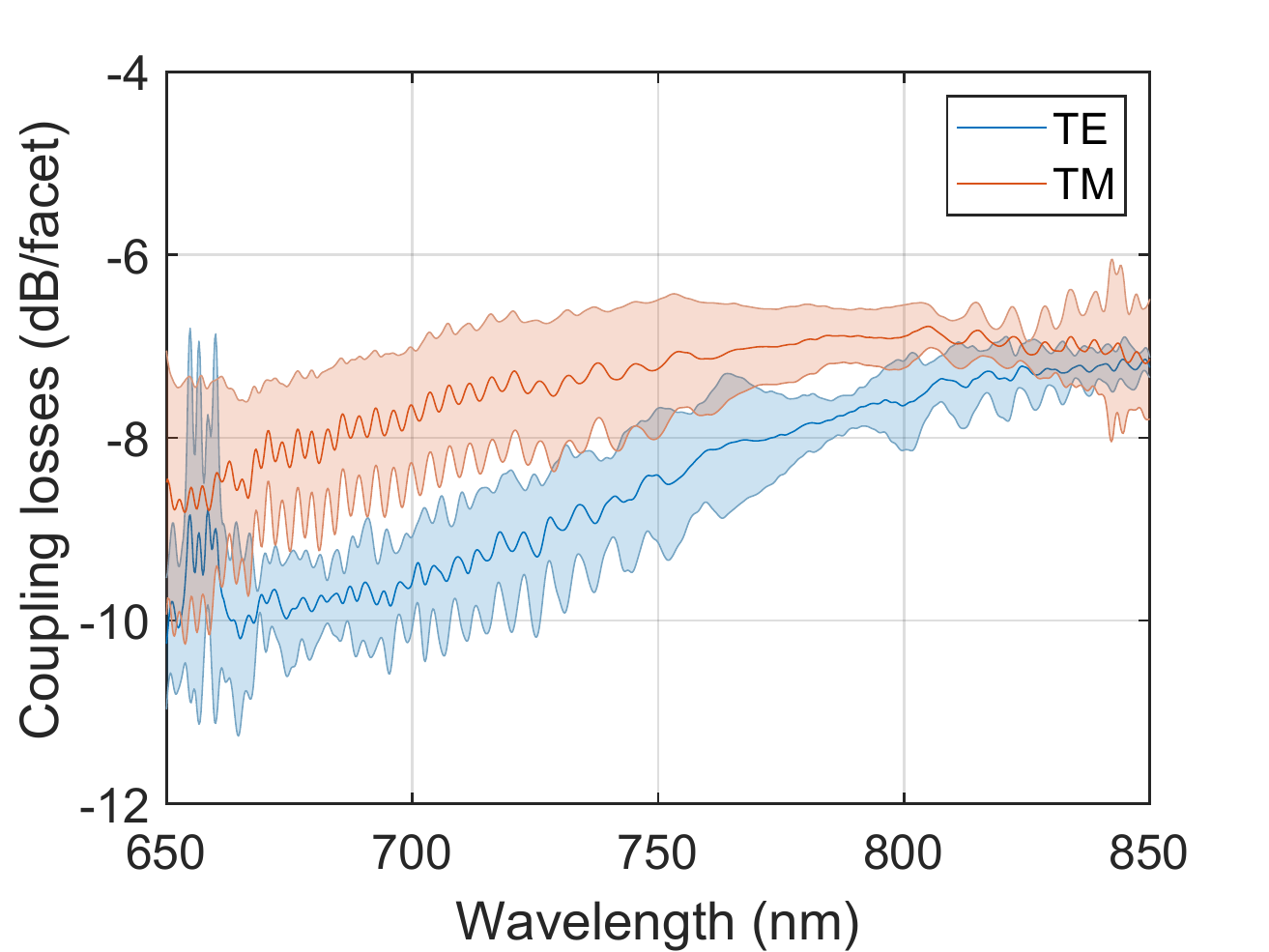}
    \end{minipage}
\caption{Measured insertion losses in the spectral region between $650-850$ nm. a) Propagation losses in TE (blue) and TM polarization (orange). b) Butt coupling losses with a waveguide edge cross section of $0.14$~$\mu$m $\times$ $3.25$~$\mu$m using a tapered fiber with a spot size of 1.25 $\mu$m for TE (blue) and TM (orange) polarized light.}
\label{prop}
\end{figure}

The first relevant figure of merit in the characterization of the performances of the fabricated waveguides is given by the insertion losses (ILs), obtained by the sum of propagation losses (PLs) and coupling losses (CLs). Performing the cut-back method \cite{keck1972spectral}, it becomes possible to accurately determine the PLs and ILs.  Fig.~\ref{prop}(a) illustrates the PLs across a range of wavelengths from 650 to 850 nm. PLs were evaluated in TE (blue line) and TM (orange line) polarizations. The highlighted band represents the uncertainty in the measurement, which was calculated by considering multiple measurements performed on three nominally-identical PICs. This approach provides a range of values that accounts for potential variations among the PICs fabricated for this study. Within the operational range around 750 nm, we observed PLs of (2.4 $\pm$ 0.2) dB/cm in TE polarization and (1.6 $\pm$ 0.2) dB/cm in TM polarization. Such an asymmetry in the PLs is caused by the higher scattering losses for TE polarization due to a more significant mode overlap with the waveguide sidewalls, whose roughness is responsible for the loss \cite{Lacey1990RadiationLF, 6750706}. We note that these results are lower than the current state of the art \cite{smith2023sin,buzaverov2023low}.

Butt coupling has been used for light injection and collection. The access waveguides of the PIC consists of an adiabatic $150 ~\mu$m-long tapered section. The waveguide width is reduced from $3.25 ~\mu$m at the facets to $0.65~\mu$m. These optimum values are the result of an optimization done via 3D FDTD simulations. After the adiabatic tapering, three Euler S-bends with an effective radius of $40~\mu$m are also inserted to radiate spurious light associated to higher-order modes. This ensures that only the fundamental mode of the two polarizations is guided in the PIC, and offsets the input and output ports of the PIC, resulting in a minimization of the collected stray light. 
CLs are shown in Fig.~\ref{prop}(b): they exhibit a value of (8.4 $\pm$ 0.8) dB per facet in TE polarization and (7.2 $\pm$ 0.8) dB per facet in TM polarization at 750 nm, which are comparable with the existing state-of-the-art \cite{buzaverov2023low}.

\section{Integrated Components}
\label{sec:intcomp}
This section focuses on the analysis of basic optical components. Achieving optimal performance of these elements is of paramount importance, as it directly impacts the overall functionality and efficiency of PICs and QPICs.

\subsection{Waveguide crossings}
\label{sec:cross}

\begin{figure}[htbp]
\centering
\includegraphics[scale=0.45]{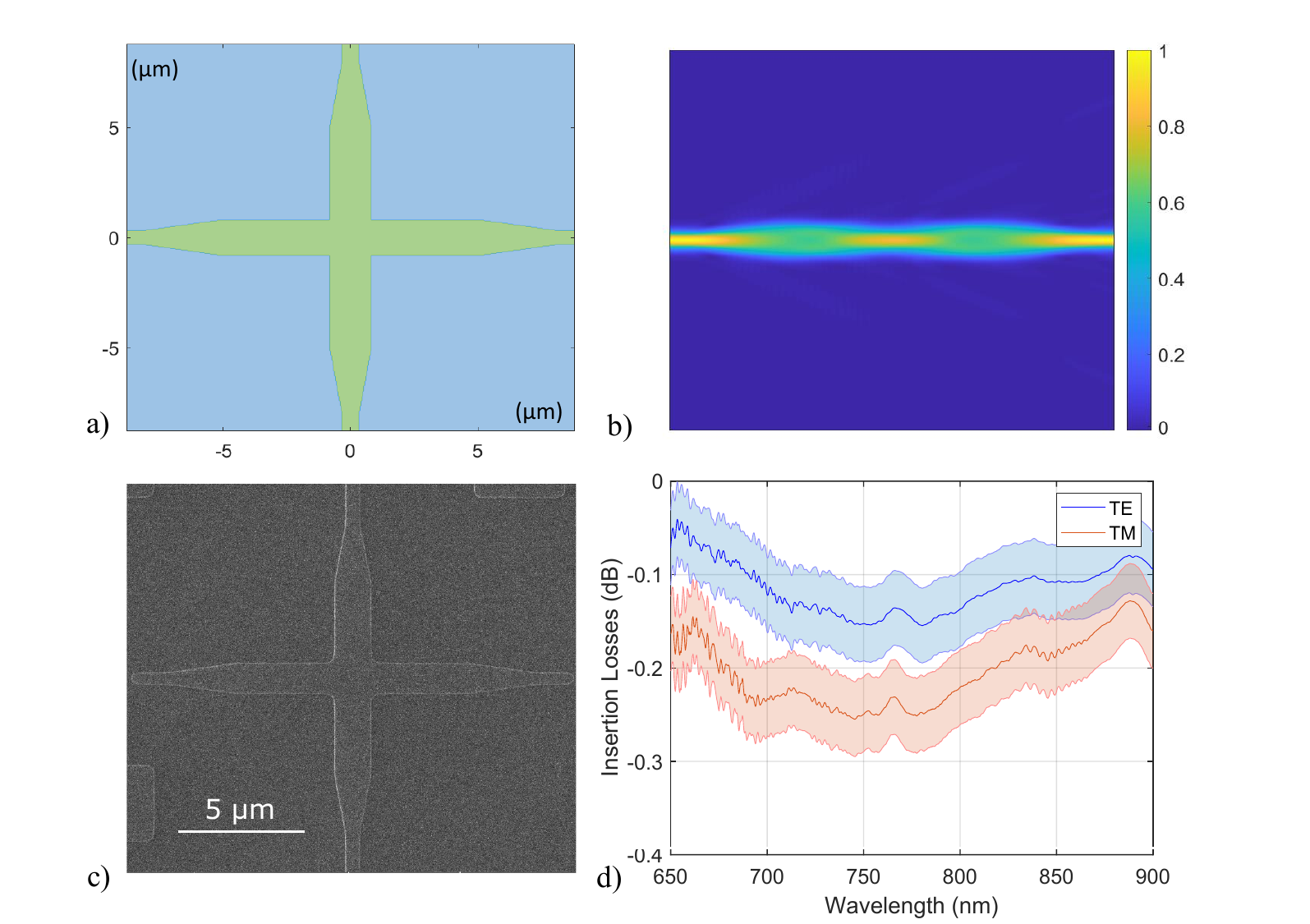}
\caption{ a) Layout of a crossing waveguide. b) Simulation of the normalized electric field for TE0 input mode in the designed crossing. c) SEM image of a crossing waveguide. d) Insertion losses in TE (blue line) and TM polarization (orange line) across the spectral region $650-850$ nm.}
\label{cross}
\end{figure}

Waveguide crossings were designed with a multimode interference-based crossing \cite{chen2006low} and optimized to negligible insertion loss and crosstalk. This design consists of a four-ports symmetric MMI with a cross geometry (Fig.~\ref{cross}(a)), where the geometrical dimensions are optimized to focus the field profile in the center of the crossing. In this way, the crosstalk between adjacent ports is minimized while the propagation to opposite ports is maximized using the self-imaging mechanism.
Fig.~\ref{cross}(a) shows the design of the crossing, whose parameters have been optimized by means of 3D FDTD simulations in the $650-850$ nm region. Fig.~\ref{cross}(b) reports the simulated electric field profile at $730$ nm for the optimized crossing where it is observed the field focused at the center of the structure. Fig.~\ref{cross}(c) shows a scanning electron microscope (SEM) image of a single crossing.

Fig.~\ref{cross}(d) shows the insertion losses for TE (blue line) and TM polarizations (orange line) of a single waveguide crossing. To assess the insertion losses accurately, the transmission of a total number of 40 cascaded crossings was measured. The uncertainty in the measurements, represented by the highlighted band, arises from multiple measurements of nominally identical crossings. In the $650-850$ nm spectral range, our crossings exhibit ILs < 0.13 dB in TE polarization and ILs < 0.25 dB in TM polarization. 
The crosstalk cannot be measured due to its low value beyond the sensitivity of the setup. Remarkably, the component shows consistent and nearly equal losses across the whole analyzed spectral region.

\subsection{MMI Beam Splitters}
\label{sec:MMI}

Beam splitters or beam couplers can be realized by MMI devices. These are composed by a certain number of input and output single-mode waveguides connected by a wide multimode waveguide. The working principle is based on self-imaging phenomenon \cite{bryngdahl1973image,ulrich1975self}. If the interference mechanism involves all the modes, the structure is called general interference MMI (G-MMI). On the other hand, if the input waveguides are positioned so that to couple to only few modes of the multimode waveguide, a different interference pattern is obtained. In particular, here we design the MMI to get pair-mode interference and we name it P-MMI. In this design the second-order, fifth-order, eighth-order, etc. modes are not excited: for every pair of excited modes, there is one which is not excited \cite{Soldanoilgrande}. Generally, G-MMIs are longer than P-MMIs and are more robust with respect to fabrication errors due to their tolerance with respect to the input waveguide position. They have superior performances, both in terms of insertion losses and bandwidth of operation. On the other hand, MMIs with restricted interference mechanisms, like P-MMIs, have a smaller footprint \cite{Soldanoilgrande}.

\begin{figure}[htbp]
\centering
\includegraphics[scale=0.28]{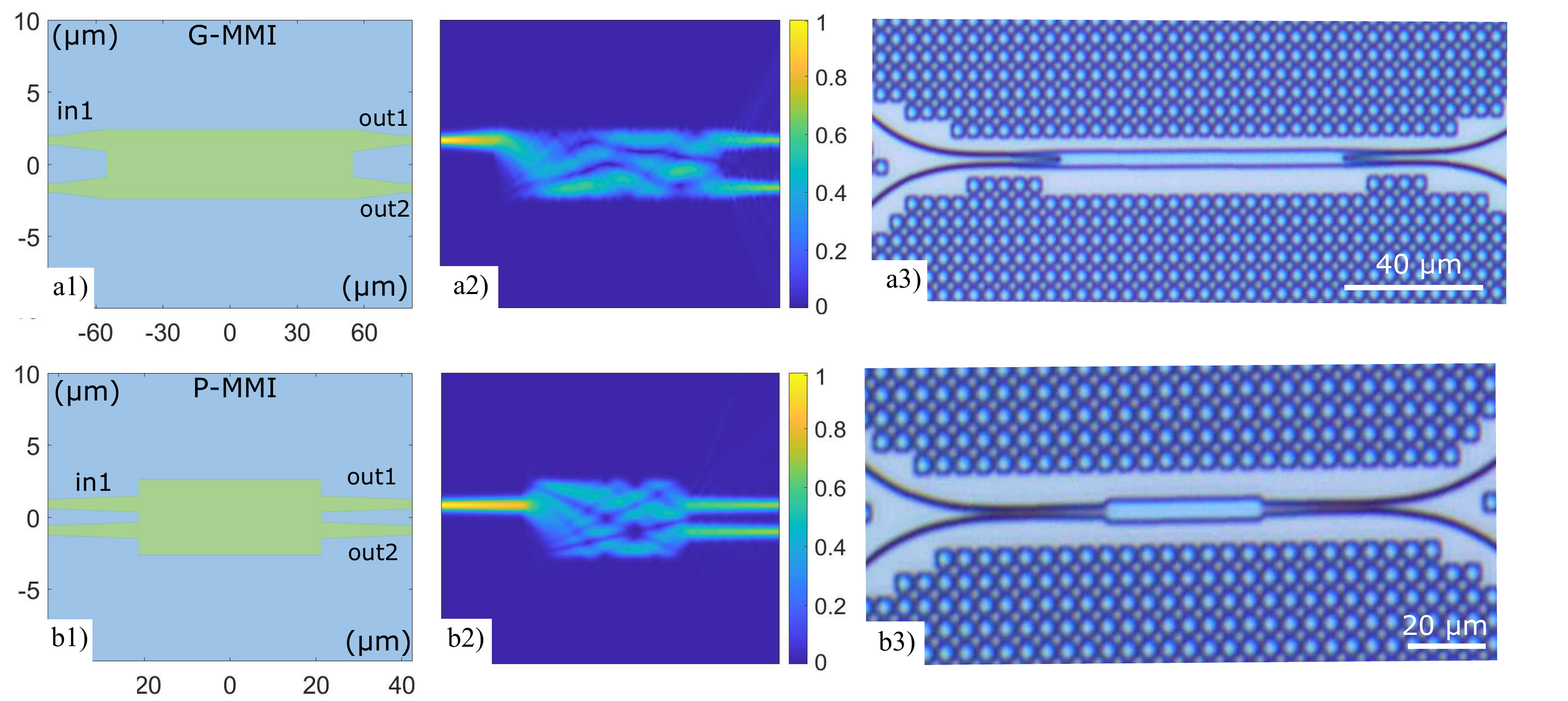}\\

\caption{a1-b1) Layouts of a 2x2 beam splitter based on G-MMI (top panel) or P-MMI (bottom panel). When used as a 1x2 beam splitter only one input port (in1) and the two output ports (out1 and out2) are used. a2-b2) FDTD simulation of the normalized electric field profile for a TE0 input mode at 730 nm. a3-b3)  SEM image of the fabricated devices.}
\label{MMI_foot_field}
\end{figure}

Figs.~\ref{MMI_foot_field}(a1-b1) show the designs of a 2x2 beam splitters based on G-MMI or P-MMI: the values of the parameters have been obtained from 3D FDTD simulations at $730$ nm for a TE0 input. Note that the length is almost three times shorter for P-MMI, and that the input and output ports of the G-MMI are set on the edge of the multimode waveguide. Figs.~\ref{MMI_foot_field}(a2-b2) report the simulated electric field profile at $730$ nm in the optimized G-MMI and P-MMI used as a 1x2 beam splitter. A different interference pattern can be observed, which arises from the different number of modes excited in the multimode waveguide. Figs.~\ref{MMI_foot_field}(a3-b3) show SEM images of a fabricated G-MMI and P-MMI.

\begin{figure}[ht!]
    \begin{minipage}{0.49\textwidth}
    \centering
    {\small a)}\includegraphics[scale=0.45]{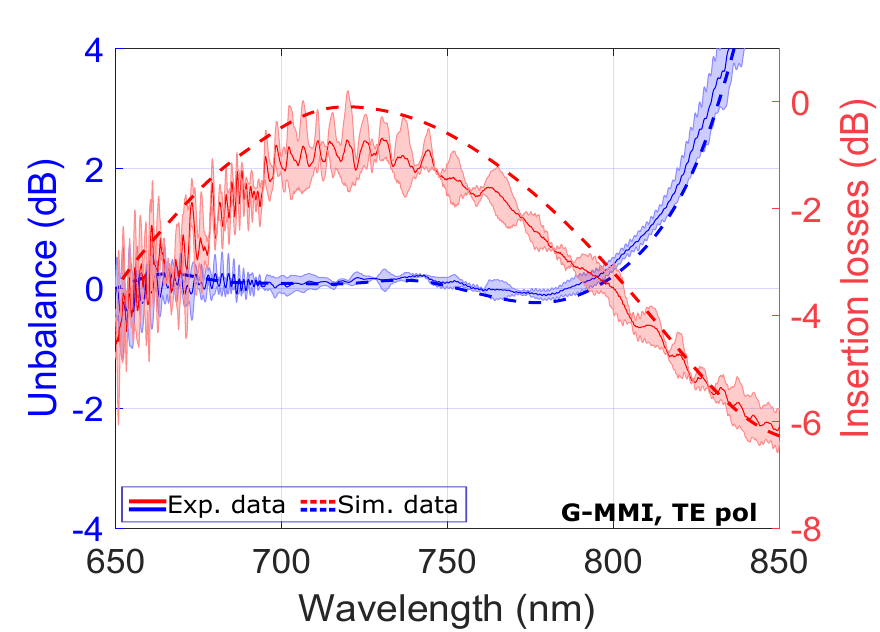}
    \end{minipage}
    \begin{minipage}{0.49\textwidth}
   \centering
    {\small b)}\includegraphics[scale=0.45]{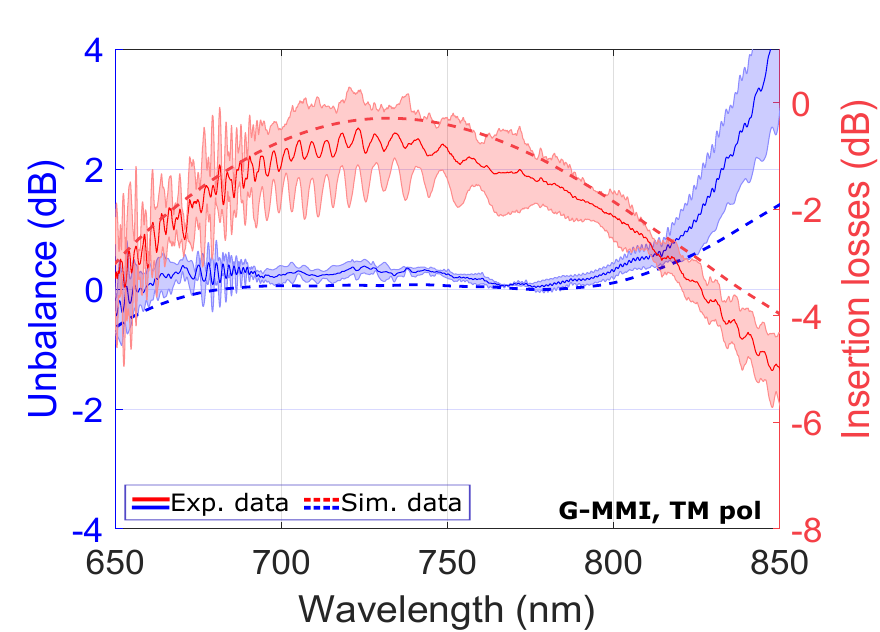}
    \end{minipage}\\
    \begin{minipage}{0.49\textwidth}
   \centering
    {\small c)}\includegraphics[scale=0.45]{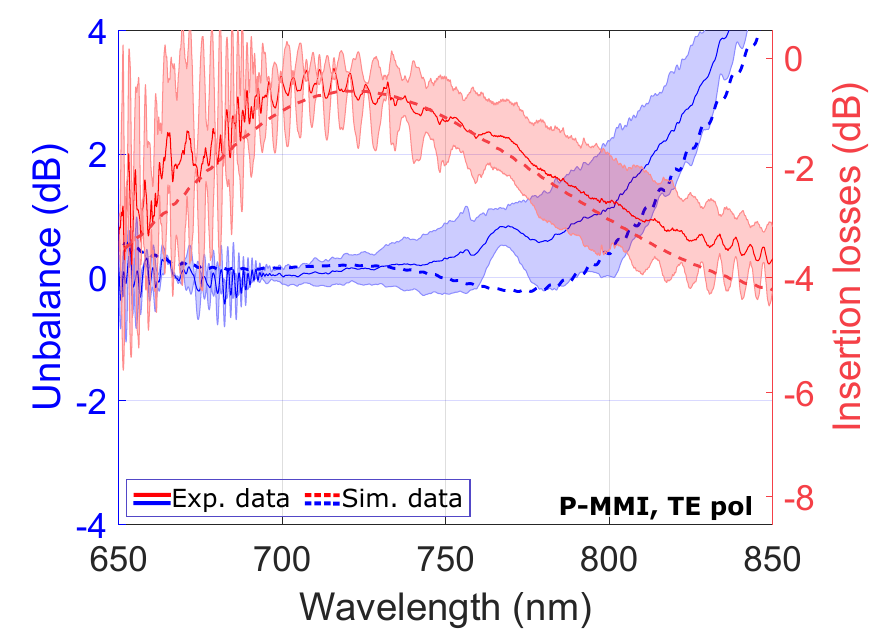}
    \end{minipage}
    \begin{minipage}{0.49\textwidth}
   \centering
    {\small d)}\includegraphics[scale=0.45]{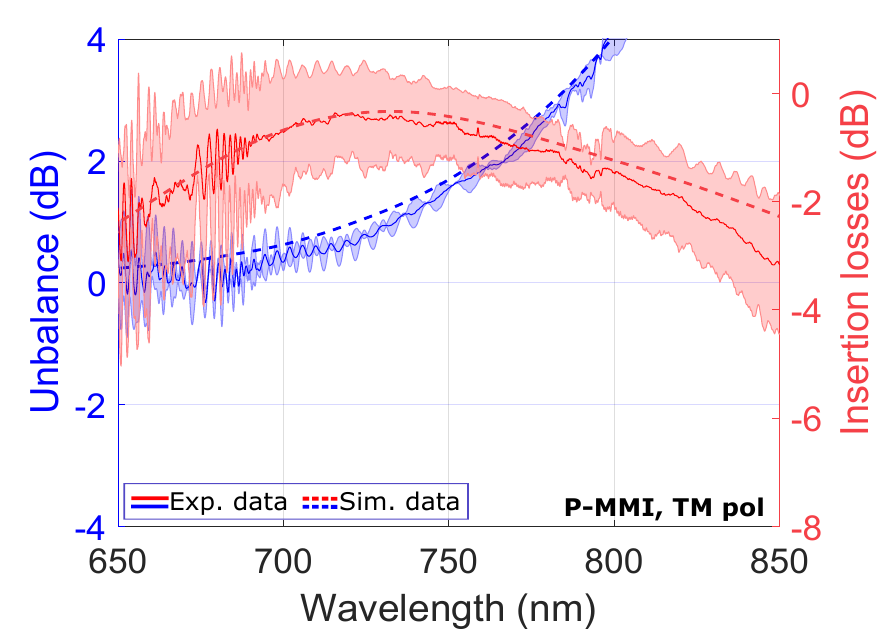}
    \end{minipage}
\caption{Insertion losses (red lines) and output unbalance (blue lines) of the G-MMI (a-b) and P-MMI (c-d) for TE and TM polarization. These are estimated by measuring the transmitted intensities from the output ports ($out_1$ and $out_2$). Then insertion losses are calculated by $10\times log(out_1+out_2)$ while the unbalance by $10\times log(out_2/out_1)$. $out$s are normalized with respect to a reference value. Lines represent experimental results, dashed lines represent simulations.}
\label{SiN_GMMI_TE}
\end{figure}

Fig.~\ref{SiN_GMMI_TE} shows the insertion losses (red lines) and output unbalance (blue lines) of 2x2 beam splitters based on G-MMI or P-MMI within a wavelength range of $650-850$ nm and for different polarizations. The uncertainties in the measurements, represented by the highlighted bands, come from multiple measurements of nominally identical MMIs. The dashed curves are the simulated spectra from 3D FDTD calculations. The insertion losses have a minimum at the design wavelength of 730 nm, with ILs < 0.5 dB for G-MMI, while this is at 720 nm for P-MMIs. For the G-MMI, the output unbalance (defined as the ratio between the transmission intensities at the two outputs) is flat over a 650-800 nm wide bandwidth with a maximum value of (0.2$\pm$ 0.2) dB for TE and (0.3$\pm$ 0.2) dB for TM polarization. For P-MMIs, an unbalance of approximately (0.7$\pm$ 0.6) dB is observed at the design wavelength for TE polarization. However, the behavior is suboptimal for TM polarization, displaying unbalance values exceeding (1.2$\pm$ 0.2) dB.
Thus, G-MMIs exhibit polarization insensitivity and demonstrate excellent outputs balance. In contrast, P-MMIs are adversely affected by polarization, leading to noticeable performance degradation in TM with respect to TE due to the fact that this component was optimized for a TE0 input mode and its selective modal excitation requirement results in a wavelength and polarization-sensitive device. Despite being less robust in terms of unbalance, P-MMIs exhibit a wider bandwidth of minimal insertion losses compared to G-MMIs.

Table~\ref{tab_mmi} provides a comparison of the 2x2 G-MMI beam splitter to the state-of-art for SiN in the ViS-NIR region. Note that the different components refer to different spectral regions.

\begin{table}[!ht]
    \centering
    \begin{tabular}{c|c|c|c|c}
        MMI ref. & Operating wavelength & Unbalance &Polarization& Insertion losses\\ \hline
         \cite{sacher2019visible} & 484.8 nm & 1.1 dB & TE & (0.19-0.47) dB \\
         \cite{ruedas2021basic} & 633 nm & 0.2 dB & TE & / \\
         This work G-MMI & 700-800 nm & 0.2-0.3 dB & TE-TM & <0.5 dB \\
         
    \end{tabular}
    \caption{Comparison of SiN MMIs in the visible (VIS) and NIR spectral region.}
    \label{tab_mmi}
\end{table}

\subsection{Mach-Zehnder Interferometers}
\label{sec:AMZI}

\begin{figure}[ht!]
    \begin{minipage}{0.49\textwidth}
    \centering
    {\small a)}\includegraphics[scale=0.45]{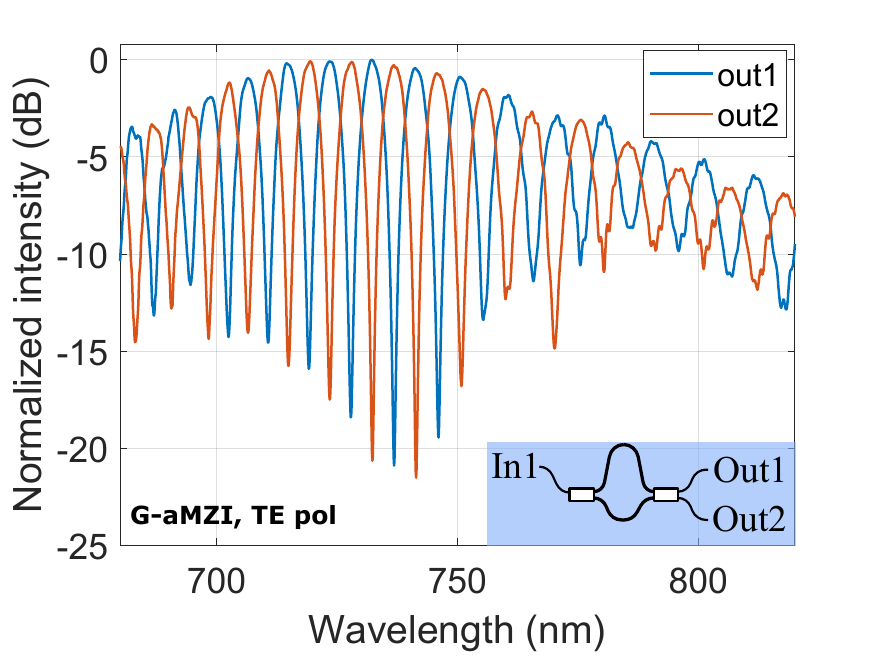}
    \end{minipage}
    \begin{minipage}{0.49\textwidth}
   \centering
    {\small b)}\includegraphics[scale=0.45]{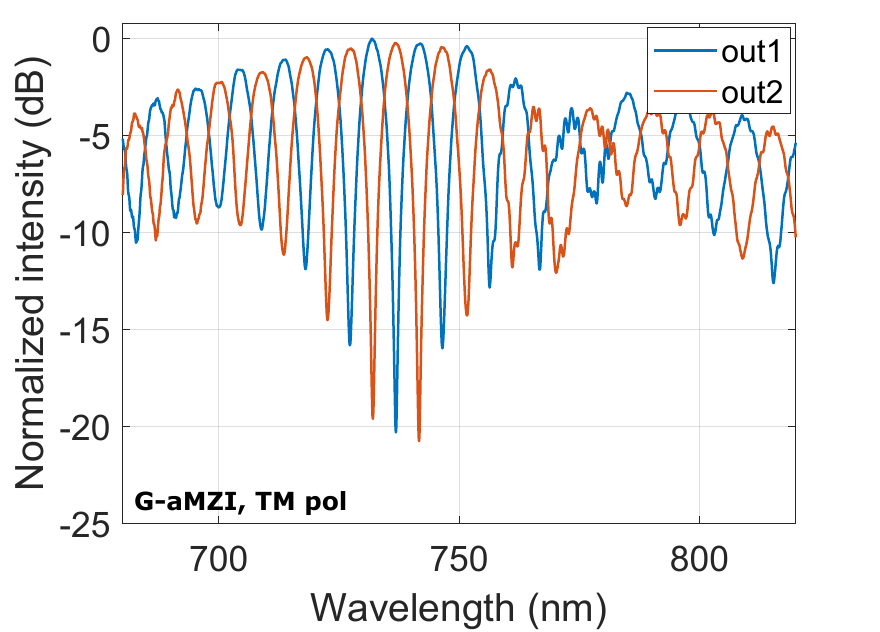}
    \end{minipage}\\
    \begin{minipage}{0.49\textwidth}
   \centering
    {\small c)}\includegraphics[scale=0.45]{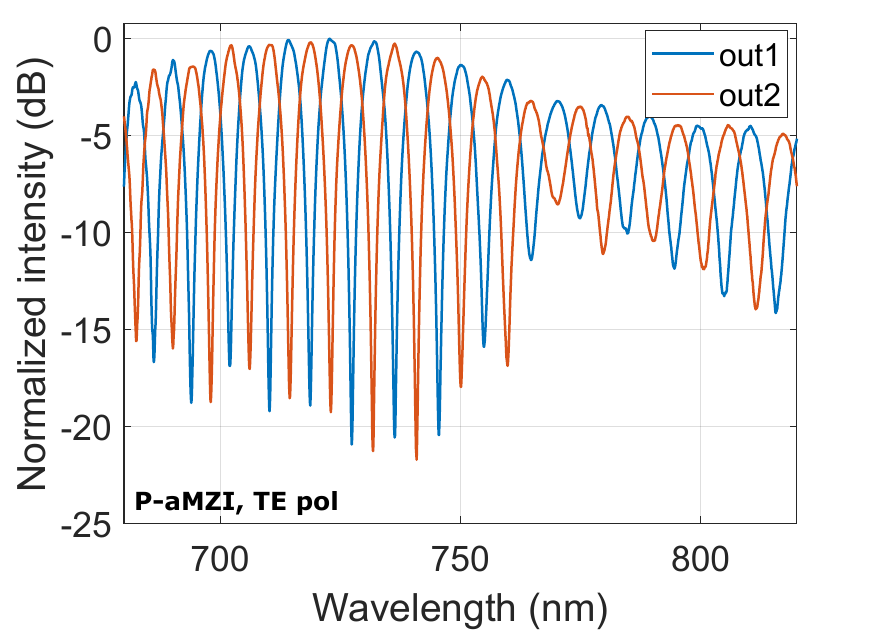}
    \end{minipage}
    \begin{minipage}{0.49\textwidth}
   \centering
    {\small d)}\includegraphics[scale=0.45]{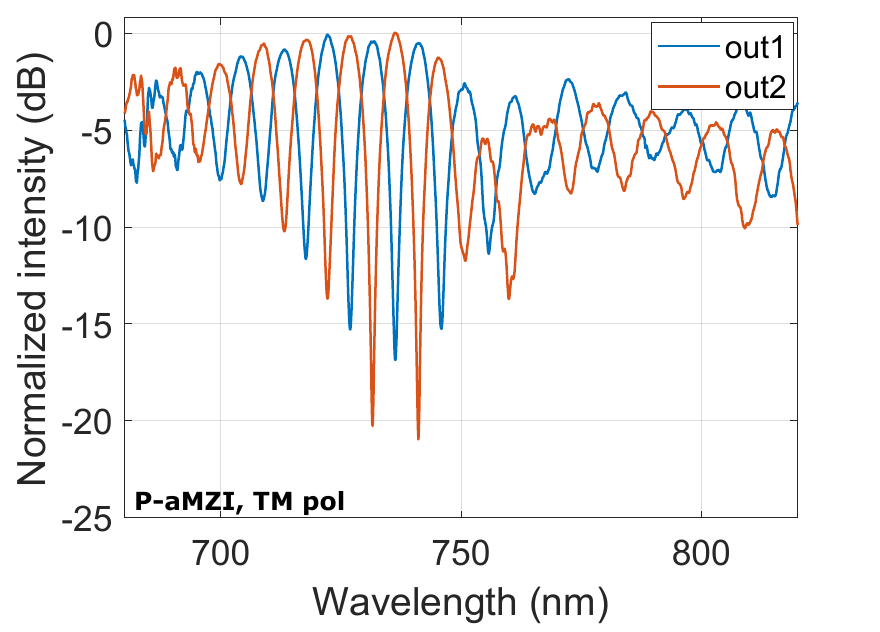}
    \end{minipage}
\caption{
Spectral transmission out of the two output ports of an aMZI based on G-MMIs in TE (a) and TM (b) polarization. Spectral transmission out of the two output ports of an aMZI based on P-MMIs in TE (c) and TM (d) polarization. The inset in (a) shows a schematic design of the aMZIs.}
\label{SiN_AMZI}
\end{figure}

From 2x2 MMIs, an integrated MZI can be realized (inset Fig.~\ref{SiN_AMZI} a). Tuning of the MZI response is usually achieved by inserting phase shifters on its arms. Therefore, an integrated MZI implements a beam splitter with controllable reflectivity that routes the optical signal in PICs or that realizes any single-qubit operation in QPICs. In fact, $m$x$m$ MZI network either in the Reck \cite{reck} or Clements \cite{clements} scheme can perform any discrete unitary operator on $m$ optical modes. Here, we consider asymmetric Mach-Zehnder Interferometers (aMZI) where the input/output 2x2 MMIs are connected by two arms of significantly different length.
Such aMZI produces an output transmission pattern characterized by a free spectral range ${\rm FSR} = \frac{\lambda^2}{n_g \cdot\delta L}$, and resonance wavelength, $\lambda_{\rm res}= \frac{n_{\rm eff} \cdot\delta L}{q}$,
where $n_{\rm eff}$ is the effective index and $n_g$ is the group index of the propagating mode in the waveguide, $q$ is an integer number and $\delta L$ is the difference between the lengths of the two aMZI arms \cite{RIZAL2018947,shamy_22}. We choose $\delta L = 30.6~\mu$m in order to have a ${\rm FSR} = 8.9$ nm, given $n_g=1.961$ at $\lambda=730$ nm. The aMZI's arms are based on four identical Euler 45$^\circ$-bends with effective radius $80~\mu$m connected by straight 1 $\mu$m wide waveguides with total length difference equal to $\delta L$: these parameters were chosen to eliminate possible loss channels. The insertion losses of the aMZIs are mainly given by the insertion losses of the MMIs, while the visibility of the interference pattern depends on the unbalance of the MMIs. This implies that the spectral bandwidth of the MZIs is intrinsically linked to the operational spectral bandwidth of the MMIs.

Fig.~\ref{SiN_AMZI} presents the spectral responses of aMZIs based on G-MMIs (G-aMZI, panels a-b) or P-MMIs (P-aMZI, panels c-d) for TE and TM polarizations, respectively.
The optical signal is input in port 1 (in1) and the transmission out of output ports 1 ($out1$, blue lines) and 2 (out 2, red lines) is measured. For the G-aMZI, insertion losses amount to (-0.8 $\pm$ 0.6) dB at 730 nm with a maximum rejection ratio of (-21.1 $\pm$ 0.4) dB for TE polarization (Fig.~\ref{SiN_AMZI}(a)). The measured FSR is (9.00 $\pm$ 0.12) nm, which is consistent with the design value. For TM polarization (Fig.~\ref{SiN_AMZI}(b)), the insertion losses in the band centered at 730 nm are (-0.7 $\pm$ 0.8) dB, with a rejection ratio of (-20.6 $\pm$ 0.6) dB. On the contrary, for the P-aMZI, Fig.~\ref{SiN_AMZI}(c) shows an insertion losses band centered at 720 nm with a minimum ILs = (-0.8 ± 0.8) dB and a rejection ratio of (-20.9 ± 0.4) dB for the chosen TE polarization. The measured FSR is (9.0± 0.3) nm. Fig.~\ref{SiN_AMZI}(d) shows that for TM polarization the insertion losses are (-0.8 $\pm$ 0.8) dB in a band consistently centered at 720 nm, with a rejection ratio of (-18.7 ± 1.2) dB. These lower performances are due to the worse behavior of P-MMIs in TM polarization, which are affected by larger insertion losses and whose unbalance impacts the rejection ratio. It is noteworthy that outside the design wavelength region, the visibility decreases, resulting in increased insertion losses and reduced rejection. We attribute this degradation of the performances to the MMI bandwidth. These data show that G-aMZIs are preferable for both the signal polarizations, while P-aMZIs can be satisfactorily used for TE polarization only. 

\subsection{High rejection filters}
One of the most critical components in QPIC is an integrated wavelength-selective optical interference notch filter with high rejection which can be used to e.g. remove high-power pump photons from low-power idler and signal photons generated in a spontaneous four-wave mixing process \cite{wang18,ba0_22,Qiang_2018, Paesani_2019, Adcock_2019}. This can be realized by using a sequence of aMZI, summing the rejection given by the single elements\cite{piekarek2017high}. The filter can be designed to operate at a specific wavelength by fixing the FSR and $\lambda_{\rm res}$ of its elements.

\begin{figure}[htbp]
\centering
\includegraphics[scale=0.47]{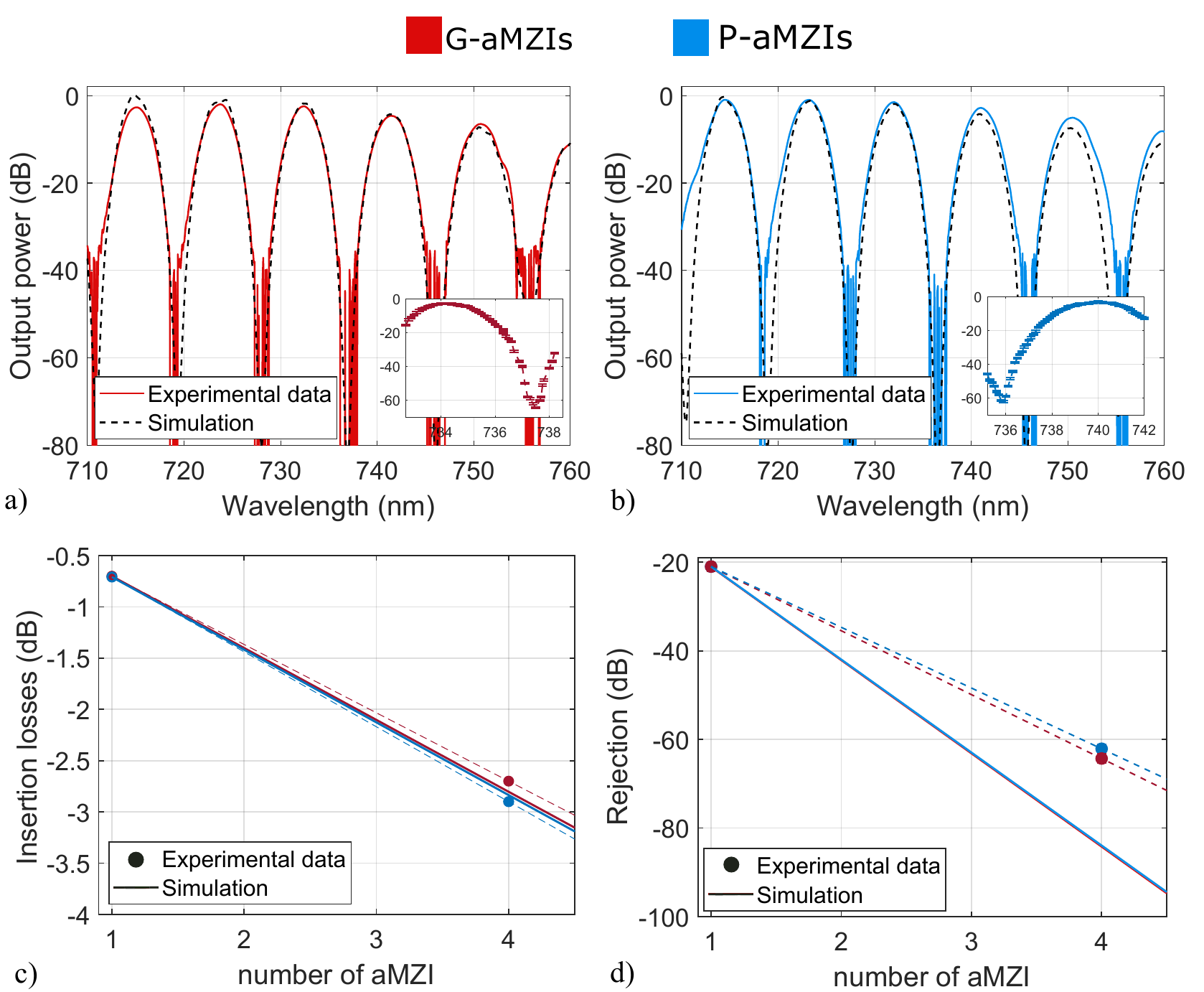}
\caption{Spectral response of a high-rejection filter based on a sequence of 4 aMZIs. (a) filter made by 4 G-aMZI. (b) filter made by 4 P-aMZI. The dashed lines refer to the simulated response. The insets show a high-resolution measurement. (c) Measured (dots) and simulated (line) insertion losses at 735 nm as a function of the number of aMZI in the sequence (red G-aMZI, blue P-aMZI). (d) Measured (dots) and simulated (line) rejection ratio at about 736 nm as a function of the number of aMZI in the sequence (red G-aMZI, blue P-aMZI).}
\label{circuit}
\end{figure}

In Fig.~\ref{circuit}, we present the results of a sequence of 4 G-aMZIs or P-aMZIs in TE polarization. The G-aMZIs and P-aMZIs sequence were measured without any trimming. Insertion losses of (-2.7 $\pm$ 0.3) dB for the G-aMZIs and (-2.9 $\pm$ 0.4) dB for the P-aMZIs are observed. The rejection ratio was measured using a tunable wavelength titanium-sapphire (Ti:Sa) laser and an optical power meter (Thorlabs-PM100USB) yielding a value of (64 ± 2) dB and (61 ± 2) dB for G-aMZIs and P-aMZIs, respectively (insets to Fig.~\ref{circuit}(a-b)). Fig.~\ref{circuit}(c) shows the insertion losses at about 735 nm for a single aMZI and for the sequence of 4 aMZI (points). Simulations are also shown as a continuous line and match the experimental results. However, when the rejection ratio is considered (Fig.~\ref{circuit}(d)), a difference is observed between the simulated and measured values. This can be attributed to the lack of active tuning of the four cascaded aMZI, which would facilitate phase synchronization, potentially enhancing the rejection. Despite this, the observed rejection ratio is high which can be attributed to the robustness of the manufacturing process of the MMIs, rendering them well-suited for this specific application. Our results cannot be compared with the literature data since there are no similar published studies in this spectral region. Nevertheless, other filter structures have been studied \cite{liu2021silicon}, such as cascaded grating-assisted contra-directional couplers (GACDC)\cite{nie2019high} or directional couplers (DC) interwoven with Bragg gratings (BG) \cite{schweikert2020design}. In the first instance, a sequence of 16 GACDC structures yielded a rejection of 68.5 dB and ILs of -5.6 dB. Our aMZI-based structures, while achieving an equivalent rejection level, exhibit significantly reduced ILs. In the second alternative filter structure, simulations show rejection of 60 dB. Table~\ref{tab_filter} summarizes this comparison.

\begin{table}[!ht]
    \centering
    \begin{tabular}{c|c|c|c|c}
        Filters & Operating wavelength & Insertion losses &Rejection& Ref.\\ \hline
        GACDC & 770-786 nm & -5.6 dB & 68.5 dB & \cite{nie2019high}  \\
         DC with BG & 785 nm & Not reported  & 60 dB (sim.) & \cite{schweikert2020design}\\
          Sequence of aMZIs & 700-800 nm & -2.7 dB & (64 $\pm$ 2) dB & This work \\
    \end{tabular}
    \caption{Comparison of SiN integrated filters in the visible (VIS) and NIR spectral region.}
    \label{tab_filter}
\end{table}

\subsection{Micro-ring resonators}

\begin{figure}[ht!]
    \centering
    {\includegraphics[scale=0.33]{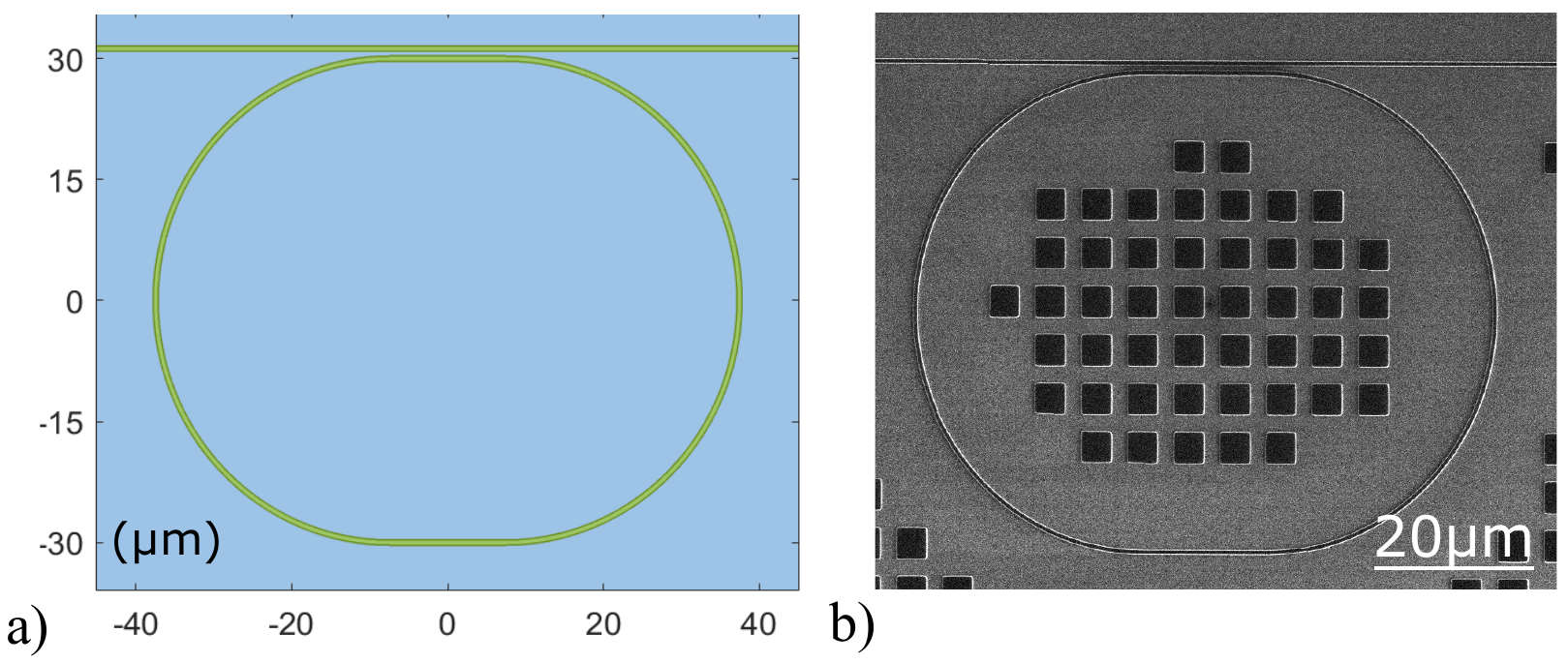}}
\caption{a) The design of the racetrack microresonator with $R=30\,\,\mu$m. b) SEM image of a microresonator.}
\label{ring_sc}
\end{figure}

Micro-ring resonators (MR) are important components in PICs given their resonant characteristics and small footprint \cite{heebner2008optical,ramiro2012fully,bogaerts2012silicon}. In QPICs, the narrow bandwidth of high-Q factor MRs can be employed to implement selective optical switches and filters, as well as for single photon processing \cite{HOM_ring, kaulfuss2023backscattering}. In addition, their resonant field enhancement can be used to enhance nonlinear optical processes in view of the generation of single photon pairs through e.g. four-wave mixing process\cite{cernansky2018complementary, Azzini2012, Engin2013}.
Fig.~\ref{ring_sc}(a) shows the design of a racetrack MR in the all-pass configuration, composed by two semicircles of radius $R=30\,\,\mu$m connected by two straight waveguides of length $L_C$, resulting in a total cavity length of $\mathcal{P} = 2\pi R + 2 L_C$. The bus waveguide runs parallel to one of the two straight arms at a distance of $600$~nm, determining the evanescent coupling strength. By using Lumerical's 2.5D FDTD Propagation Method, we simulate such design with different coupling lengths $L_C$ in the spectral range $650-850$ nm. Fig.~\ref{ring_sc}(b) reports the SEM image of one of the fabricated structures.

\begin{figure}[htbp]
\centering
{\small a)}\includegraphics[scale=0.47]{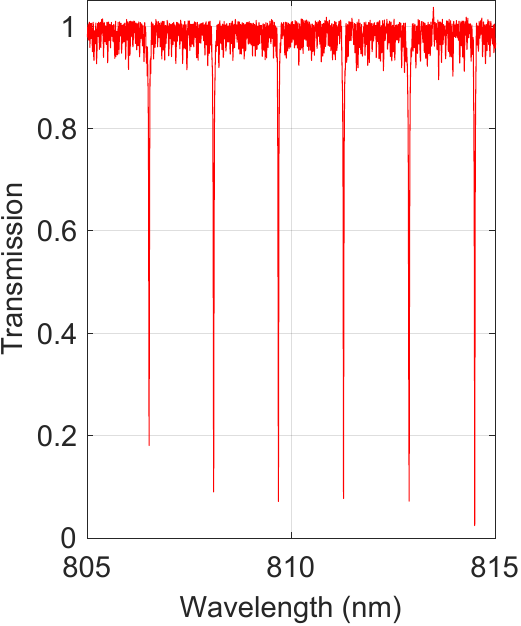}{\small b)}\includegraphics[scale=0.47]{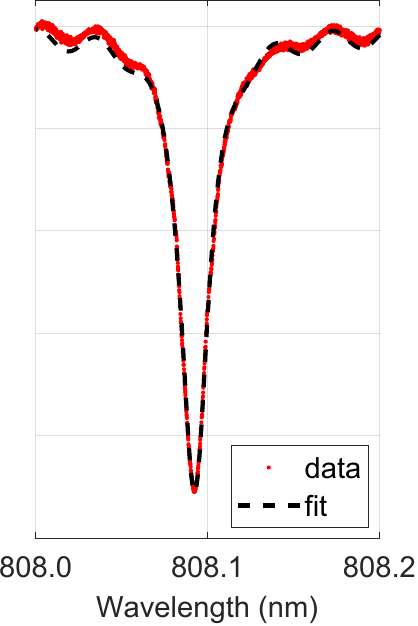}
{\small c)}\includegraphics[scale=0.47]{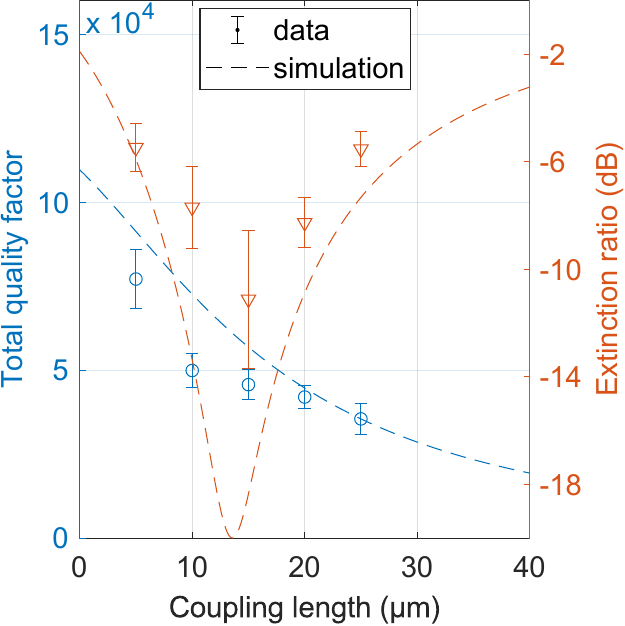}
\caption{Spectral characterization of racetrack microresonators with $R=30\,\,\mu$m. a) Transmission spectrum. b) Fit of a single resonance. c) Total Q-factor (light blue) and Extinction ratio (orange) as a function of the coupling length $L_C$. Points represent experimental data. Dashed lines show the simulation results, obtained by fixing the intrinsic quality factor to the simulated value $Q_i = 1.2\times 10^5$. The minimum of the simulated extinction ratio is set to -20 dB for visualization purposes.}
\label{Fig_Ring}
\end{figure}

To find the critical-coupling working point of our MR, a series of racetrack MRs with different $L_C$, from 5 $\mu$m to 25 $\mu m$, was fabricated. Notably, the variation of $L_C$ changes the expected Free-Spectral-Range of the MR, ${\rm FSR}(\lambda) = \frac{\lambda^2}{n_g \cdot\mathcal{P}}$, but not its intrinsic quality factor, which depends only on the propagation loss $Q_i = \frac{\pi\cdot n_g}{\lambda\cdot {\rm PLs}}$ \cite{niehusmann2004ultrahigh}. Then, by analyzing MRs with different tailored $L_C$ it is possible to determine the critical-coupling regime as the point at which the extinction ratio ER shows a minimum and the total quality factor $Q^{-1}_{tot} = Q^{-1}_i + Q^{-1}_{L_C}$ becomes half the value of the intrinsic one \cite{atvars2021analytical,niehusmann2004ultrahigh,bogaerts2012silicon}.
Since the spectral resolution of the setup presented in Fig.~\ref{setup}(a) is not suited to measure high-Q factor MR, a different measurement setup was implemented using a tunable ECDL source (Sacher Lion) and an optical wavelength meter (HighFinesse WS6) with $1$~pm accuracy. A series of MRs was probed in the TE polarization in $805-815$~nm range, obtaining normalized transmission spectra as the one shown in Fig.~\ref{Fig_Ring}(a). Fig.~\ref{Fig_Ring}(b) details a single resonance and its fit with a single Lorentzian. Fig.~\ref{Fig_Ring}(c) shows the coupling length dependence of the Q-factor (light blue points) and of the ER (red points) compared with the simulations (dashed lines). A nearly critical coupling regime is observed for $L_C^{crit} = 15.0~\mu$m with a measured $Q_{tot}^{crit} = (4.5 \pm 0.2)\times 10^4$ and ER= (-11 $\pm$ 3) dB. These values match the simulated intrinsic $Q_i = (1.20 \pm 0.05)\times 10^5$, which corresponds to a value for the propagation loss inside the MR of ${\rm PLs}_{ring}=2.6\pm0.1$ dB/cm, in good agreement with the PLs measured directly at $810$ nm in straight waveguides.

\section{Conclusion}
\label{sec:conc}
This work demonstrates that SiN is a suitable material to develop QPICs in the visible-to-near-infrared region. We have designed, fabricated, and tested different basic building blocks integrated in SiN PIC. All the devices we have described exhibit state-of-the-art performances, characterized by low insertion losses and high efficiency. The multimode interferometers (MMIs), exhibiting a spectral unbalance below 0.2 dB across a bandwidth exceeding 100 nm, have demonstrated remarkable versatility. This device is suited for Mach-Zehnder interferometers (MZIs). These latter components have demonstrated their efficacy in the implementation of specific wavelength selectors and integrated filters, enabling rejection levels of approximately 21 dB with a single device and over 60 dB in a cascade of four fully passive devices. Our micro-ring resonators have achieved a high-quality factor $Q_{tot}^{crit}$ of 4.5 $\times$ 10$^4$ while maintaining compact dimensions. This combination of a high Q factor and a significant non-linear coefficient $n_2$ further stress the potential of SiN for integrated quantum applications. We are currently working to develop quantum photon sources and single photon detectors on this platform to close the set of building blocks needed to realize a complete QPIC in SiN around 750 nm.

\section*{Acknowledgements}
This work was supported by EC within the Horizon 2020 project EPIQUS (grant 899368) and by Q@TN, the joint lab between University of Trento, FBK- Fondazione Bruno Kessler, INFN-National Institute for Nuclear Physics and CNR-National Research Council and financed by PAT.

\medskip

\bibliography{Optica-template}

\begin{thebibliography}{10}
\newcommand{\enquote}[1]{``#1''}

\bibitem{degen2017quantum}
C.~L. Degen, F.~Reinhard, and P.~Cappellaro, \enquote{Quantum sensing,}
  {\protect\JournalTitle{Reviews of modern physics}} \textbf{89}, 035002
  (2017).

\bibitem{gisin2007quantum}
N.~Gisin and R.~Thew, \enquote{Quantum communication,}
  {\protect\JournalTitle{Nature photonics}} \textbf{1}, 165--171 (2007).

\bibitem{DiVinc}
D.~P. DiVincenzo, \enquote{Quantum computation,}
  {\protect\JournalTitle{Science}} \textbf{270}, 255--261 (1995).

\bibitem{nielsen2001quantum}
M.~A. Nielsen and I.~L. Chuang, \enquote{Quantum computation and quantum
  information,} {\protect\JournalTitle{Phys. Today}} \textbf{54}, 60 (2001).

\bibitem{arute2019quantum}
F.~Arute, K.~Arya, R.~Babbush, D.~Bacon, J.~C. Bardin, R.~Barends, R.~Biswas,
  S.~Boixo, F.~G. Brandao, and D.~A. e.~a. Buell, \enquote{Quantum supremacy
  using a programmable superconducting processor,}
  {\protect\JournalTitle{Nature}} \textbf{574}, 505--510 (2019).

\bibitem{raussendorf2001a}
R.~Raussendorf and H.~J. Briegel, \enquote{A one-way quantum computer,}
  {\protect\JournalTitle{Phys. Rev. Lett.}} \textbf{86}, 5188--5191 (2001).

\bibitem{briege2009measurement-based}
H.~Briegel, D.~Browne, and M.~Van~den Nest, \enquote{Measurement-based quantum
  computation,} {\protect\JournalTitle{Nature Physics}} \textbf{5}, 19--26
  (2009).

\bibitem{bombin2021modular}
H.~Bombin, I.~H. Kim, D.~Litinski, and et~al., \enquote{Interleaving: Modular
  architectures for fault-tolerant photonic quantum computing,}
  {\protect\JournalTitle{arXiv preprint arXiv:2103.08612}}  (2021).

\bibitem{rudolph2017optimistic}
T.~Rudolph, \enquote{{Why I am optimistic about the silicon-photonic route to
  quantum computing},} {\protect\JournalTitle{APL photonics}} \textbf{2}
  (2017).

\bibitem{wang18}
J.~Wang, S.~Paesani, Y.~Ding, R.~Santagati, P.~Skrzypczyk, A.~Salavrakos,
  J.~Tura, R.~Augusiak, L.~Mančinska, D.~Bacco, D.~Bonneau, J.~W. Silverstone,
  Q.~Gong, A.~Acín, K.~Rottwitt, L.~K. Oxenløwe, J.~L. O’Brien, A.~Laing,
  and M.~G. Thompson, \enquote{Multidimensional quantum entanglement with
  large-scale integrated optics,} {\protect\JournalTitle{Science}}
  \textbf{360}, 285--291 (2018).

\bibitem{reimer19}
C.~Reimer, S.~Sciara, P.~Roztocki, M.~Islam, L.~Cort{\'e}s, Y.~Zhang,
  B.~Fischer, S.~Loranger, R.~Kashyap, A.~Cino, S.~Chu, B.~Little, D.~Moss,
  L.~Caspani, W.~Munro, J.~Aza{\~n}a, M.~Kues, and R.~Morandotti,
  \enquote{High-dimensional one-way quantum processing implemented on d-level
  cluster states,} {\protect\JournalTitle{Nature Physics}} \textbf{15},
  148--153 (2019).

\bibitem{zhong20}
H.-S. Zhong, H.~Wang, Y.-H. Deng, M.-C. Chen, L.-C. Peng, Y.-H. Luo, J.~Qin,
  D.~Wu, X.~Ding, Y.~Hu, P.~Hu, X.-Y. Yang, W.-J. Zhang, H.~Li, Y.~Li,
  X.~Jiang, L.~Gan, G.~Yang, L.~You, Z.~Wang, L.~Li, N.-L. Liu, C.-Y. Lu, and
  J.-W. Pan, \enquote{Quantum computational advantage using photons,}
  {\protect\JournalTitle{Science}} \textbf{370}, 1460--1463 (2020).

\bibitem{viglia2021error-protected}
C.~Vigliar, S.~Paesani, Y.~Ding, and et~al., \enquote{Error-protected qubits in
  a silicon photonic chip,} {\protect\JournalTitle{Nature Physics}}
  \textbf{17}, 1173--1143 (2021).

\bibitem{boson_sampling}
S.~Aaronson and A.~Arkhipov, \enquote{The computational complexity of linear
  optics,} {\protect\JournalTitle{arXiv}}  (2010).

\bibitem{boson_sampling_review}
D.~J. Brod, E.~F. Galv{\~a}o, A.~Crespi, R.~Osellame, N.~Spagnolo, and
  F.~Sciarrino, \enquote{{Photonic implementation of boson sampling: a
  review},} {\protect\JournalTitle{Advanced Photonics}} \textbf{1}, 034001
  (2019).

\bibitem{ikeda2008thermal}
K.~Ikeda, R.~E. Saperstein, N.~Alic, and Y.~Fainman, \enquote{{Thermal and Kerr
  nonlinear properties of plasma-deposited silicon nitride/silicon dioxide
  waveguides},} {\protect\JournalTitle{Optics express}} \textbf{16},
  12987--12994 (2008).

\bibitem{cernansky2018complementary}
R.~Cernansky, F.~Martini, and A.~Politi, \enquote{Complementary metal-oxide
  semiconductor compatible source of single photons at near-visible
  wavelengths,} {\protect\JournalTitle{Optics Letters}} \textbf{43}, 855--858
  (2018).

\bibitem{8472140}
D.~J. Blumenthal, R.~Heideman, D.~Geuzebroek, A.~Leinse, and C.~Roeloffzen,
  \enquote{Silicon nitride in silicon photonics,}
  {\protect\JournalTitle{{Proceedings of the IEEE}}} \textbf{106}, 2209--2231
  (2018).

\bibitem{piccoli2022silicon}
G.~Piccoli, M.~Sanna, M.~Borghi, L.~Pavesi, and M.~Ghulinyan, \enquote{{Silicon
  oxynitride platform for linear and nonlinear photonics at NIR wavelengths},}
  {\protect\JournalTitle{Optical Materials Express}} \textbf{12}, 3551--3562
  (2022).

\bibitem{leone2023generation}
N.~Leone, S.~Azzini, S.~Mazzucchi, V.~Moretti, M.~Sanna, M.~Borghi, G.~Piccoli,
  M.~Bernard, M.~Ghulinyan, and L.~Pavesi, \enquote{{Generation of
  quantum-certified random numbers using on-chip path-entangled single photons
  from an LED},} {\protect\JournalTitle{Photon. Res.}} \textbf{11}, 1484--1499
  (2023).

\bibitem{bres2023supercontinuum}
C.-S. Br{\`e}s, A.~Della~Torre, D.~Grassani, V.~Brasch, C.~Grillet, and
  C.~Monat, \enquote{Supercontinuum in integrated photonics: generation,
  applications, challenges, and perspectives,}
  {\protect\JournalTitle{Nanophotonics}} \textbf{12}, 1199--1244 (2023).

\bibitem{bernard2021top}
M.~Bernard, F.~Acerbi, G.~Paternoster, G.~Piccoli, L.~Gemma, D.~Brunelli,
  A.~Gola, G.~Pucker, L.~Pancheri, and M.~Ghulinyan, \enquote{Top-down
  convergence of near-infrared photonics with silicon substrate-integrated
  electronics,} {\protect\JournalTitle{Optica}} \textbf{8}, 1363--1364 (2021).

\bibitem{shan2014single}
G.-C. Shan, Z.-Q. Yin, C.~H. Shek, and W.~Huang, \enquote{Single photon sources
  with single semiconductor quantum dots,} {\protect\JournalTitle{Frontiers of
  Physics}} \textbf{9}, 170--193 (2014).

\bibitem{PhysRevB.72.195332}
A.~Malko, D.~Y. Oberli, M.~H. Baier, E.~Pelucchi, F.~Michelini, K.~F. Karlsson,
  M.-A. Dupertuis, and E.~Kapon, \enquote{Single-photon emission from pyramidal
  quantum dots: The impact of hole thermalization on photon emission
  statistics,} {\protect\JournalTitle{Phys. Rev. B}} \textbf{72}, 195332
  (2005).

\bibitem{malko2006optimization}
A.~Malko, M.~Baier, K.~Karlsson, E.~Pelucchi, D.~Oberli, and E.~Kapon,
  \enquote{Optimization of the efficiency of single-photon sources based on
  quantum dots under optical excitation,} {\protect\JournalTitle{Applied
  Physics Letters}} \textbf{88} (2006).

\bibitem{kimura2005single}
S.~Kimura, H.~Kumano, M.~Endo, I.~Suemune, T.~Yokoi, H.~Sasakura, S.~Adachi,
  S.~Muto, H.~Song, and S.~e.~a. Hirose, \enquote{{Single-photon generation
  from InAlAs single quantum dot},} {\protect\JournalTitle{physica status
  solidi (c)}} \textbf{2}, 3833--3837 (2005).

\bibitem{kiravittaya2006ordered}
S.~Kiravittaya, M.~Benyoucef, R.~Zapf-Gottwick, A.~Rastelli, and O.~Schmidt,
  \enquote{{Ordered GaAs quantum dot arrays on GaAs (001): Single photon
  emission and fine structure splitting},} {\protect\JournalTitle{Applied
  physics letters}} \textbf{89} (2006).

\bibitem{keck1972spectral}
D.~Keck and A.~Tynes, \enquote{Spectral response of low-loss optical
  waveguides,} {\protect\JournalTitle{Applied optics}} \textbf{11}, 1502--1506
  (1972).

\bibitem{Lacey1990RadiationLF}
J.~Lacey and F.~Payne, \enquote{Radiation loss from planar waveguides with
  random wall imperfections,} {\protect\JournalTitle{IEE Proceedings J
  (Optoelectronics)}} \textbf{137}, 282--289 (1990).

\bibitem{6750706}
C.~Qiu, Z.~Sheng, H.~Li, W.~Liu, L.~Li, A.~Pang, A.~Wu, X.~Wang, S.~Zou, and
  F.~Gan, \enquote{Fabrication, characterization and loss analysis of silicon
  nanowaveguides,} {\protect\JournalTitle{Journal of Lightwave Technology}}
  \textbf{32}, 2303--2307 (2014).

\bibitem{smith2023sin}
J.~A. Smith, H.~Francis, G.~Navickaite, and M.~J. Strain, \enquote{{SiN foundry
  platform for high performance visible light integrated photonics},}
  {\protect\JournalTitle{Optical Materials Express}} \textbf{13}, 458--468
  (2023).

\bibitem{buzaverov2023low}
K.~A. Buzaverov, A.~S. Baburin, E.~V. Sergeev, S.~S. Avdeev, E.~S. Lotkov,
  M.~Andronik, V.~E. Stukalova, D.~A. Baklykov, I.~V. Dyakonov, and N.~N. e.~a.
  Skryabin, \enquote{{Low-loss silicon nitride photonic ICs for near-infrared
  wavelength bandwidth},} {\protect\JournalTitle{Optics Express}} \textbf{31},
  16227--16242 (2023).

\bibitem{chen2006low}
H.~Chen and A.~W. Poon, \enquote{Low-loss multimode-interference-based
  crossings for silicon wire waveguides,} {\protect\JournalTitle{IEEE photonics
  technology letters}} \textbf{18}, 2260--2262 (2006).

\bibitem{bryngdahl1973image}
O.~Bryngdahl, \enquote{Image formation using self-imaging techniques,}
  {\protect\JournalTitle{JOSA}} \textbf{63}, 416--419 (1973).

\bibitem{ulrich1975self}
R.~Ulrich and G.~Ankele, \enquote{Self-imaging in homogeneous planar optical
  waveguides,} {\protect\JournalTitle{Applied Physics Letters}} \textbf{27},
  337--339 (1975).

\bibitem{Soldanoilgrande}
L.~Soldano and E.~C.~M. Pennings, \enquote{Optical multi-mode interference
  devices based on self-imaging: principles and applications,}
  {\protect\JournalTitle{Journal of Lightwave Technology}} \textbf{13},
  615--627 (1995).

\bibitem{sacher2019visible}
W.~D. Sacher, X.~Luo, Y.~Yang, F.-D. Chen, T.~Lordello, J.~C. Mak, X.~Liu,
  T.~Hu, T.~Xue, and P.~G.-Q. e.~a. Lo, \enquote{Visible-light silicon nitride
  waveguide devices and implantable neurophotonic probes on thinned 200 mm
  silicon wafers,} {\protect\JournalTitle{Optics express}} \textbf{27},
  37400--37418 (2019).

\bibitem{ruedas2021basic}
J.~F. Ruedas, J.~Sabek, T.~D. Bucio, F.~Y. Gardes, and C.~D. Horna,
  \enquote{{Basic building blocks development for a SiN platform in the visible
  range},} {\protect\JournalTitle{{2021 IEEE 17th International Conference on
  Group IV Photonics (GFP)}}} pp. 1--2 (2021).

\bibitem{reck}
M.~Reck, A.~Zeilinger, H.~J. Bernstein, and P.~Bertani, \enquote{Experimental
  realization of any discrete unitary operator,} {\protect\JournalTitle{Phys.
  Rev. Lett.}} \textbf{73}, 58--61 (1994).

\bibitem{clements}
W.~R. Clements, P.~C. Humphreys, B.~J. Metcalf, W.~S. Kolthammer, and I.~A.
  Walmsley, \enquote{Optimal design for universal multiport interferometers,}
  {\protect\JournalTitle{Optica}} \textbf{3}, 1460--1465 (2016).

\bibitem{RIZAL2018947}
C.~Rizal and B.~Niraula, \enquote{{Compact Si-based asymmetric MZI waveguide on
  SOI as a thermo-optical switch},} {\protect\JournalTitle{Optics
  Communications}} \textbf{410}, 947--955 (2018).

\bibitem{shamy_22}
e.~a. Raghi S. El~Shamy, \enquote{Modelling, characterization, and applications
  of silicon on insulator loop terminated asymmetric mach zehnder
  interferometer,} {\protect\JournalTitle{Scientific Reports}} \textbf{12}
  (2022).

\bibitem{ba0_22}
J.~Bao, Z.~Fu, T.~Pramanik, J.~Mao, Y.~Chi, Y.~Cao, C.~Zhai, Y.~Mao, T.~Dai,
  X.~Chen, X.~Jia, L.~Zhao, Y.~Zheng, B.~Tang, Z.~Li, J.~Luo, W.~Wang, Y.~Yang,
  Y.~Peng, D.~Liu, D.~Dai, Q.~He, A.~Muthali, L.~Oxenl{\o}we, C.~Vigliar,
  S.~Paesani, H.~Hou, R.~Santagati, J.~Silverstone, A.~Laing, M.~Thompson,
  J.~O{\textquoteright}Brien, Y.~Ding, Q.~Gong, and J.~Wang,
  \enquote{Very-large-scale integrated quantum graph photonics,}
  {\protect\JournalTitle{Nature Photonics}}  (2023). Publisher Copyright:
  {\textcopyright} 2023, The Author(s).

\bibitem{Qiang_2018}
X.~Qiang, X.~Zhou, J.~Wang, C.~M. Wilkes, T.~Loke, S.~O'Gara, L.~Kling, G.~D.
  Marshall, R.~Santagati, T.~C. Ralph, J.~B. Wang, J.~L. O'Brien, M.~G.
  Thompson, and J.~C.~F. Matthews, \enquote{Large-scale silicon quantum
  photonics implementing arbitrary two-qubit processing,}
  {\protect\JournalTitle{Nature Photonics}} \textbf{12}, 534--539 (2018).

\bibitem{Paesani_2019}
S.~Paesani, Y.~Ding, R.~Santagati, L.~Chakhmakhchyan, C.~Vigliar, K.~Rottwitt,
  L.~K. Oxenl{\o}we, J.~Wang, M.~G. Thompson, and A.~Laing, \enquote{Generation
  and sampling of quantum states of light in a silicon chip,}
  {\protect\JournalTitle{Nature Physics}} \textbf{15}, 925--929 (2019).

\bibitem{Adcock_2019}
J.~C. Adcock, C.~Vigliar, R.~Santagati, J.~W. Silverstone, and M.~G. Thompson,
  \enquote{Programmable four-photon graph states on a silicon chip,}
  {\protect\JournalTitle{Nature Communications}} \textbf{10} (2019).

\bibitem{piekarek2017high}
M.~Piekarek, D.~Bonneau, S.~Miki, T.~Yamashita, M.~Fujiwara, M.~Sasaki,
  H.~Terai, M.~G. Tanner, C.~M. Natarajan, and R.~H. e.~a. Hadfield,
  \enquote{High-extinction ratio integrated photonic filters for silicon
  quantum photonics,} {\protect\JournalTitle{Optics letters}} \textbf{42},
  815--818 (2017).

\bibitem{liu2021silicon}
D.~Liu, H.~Xu, Y.~Tan, Y.~Shi, and D.~Dai, \enquote{Silicon photonic filters,}
  {\protect\JournalTitle{Microwave and Optical Technology Letters}}
  \textbf{63}, 2252--2268 (2021).

\bibitem{nie2019high}
X.~Nie, N.~Turk, Y.~Li, Z.~Liu, and R.~Baets, \enquote{High extinction ratio
  on-chip pump-rejection filter based on cascaded grating-assisted
  contra-directional couplers in silicon nitride rib waveguides,}
  {\protect\JournalTitle{Optics letters}} \textbf{44}, 2310--2313 (2019).

\bibitem{schweikert2020design}
C.~Schweikert, N.~Hoppe, R.~Elster, T.~F{\"o}hn, W.~Vogel, and M.~Berroth,
  \enquote{Design of a broadband integrated notch filter in silicon nitride,}
  {\protect\JournalTitle{2020 International Conference on Numerical Simulation
  of Optoelectronic Devices (NUSOD)}} pp. 89--90 (2020).

\bibitem{heebner2008optical}
J.~Heebner, R.~Grover, and T.~Ibrahim, \enquote{Optical microresonator theory,}
  {\protect\JournalTitle{Springer}}  (2008).

\bibitem{ramiro2012fully}
F.~Ramiro-Manzano, N.~Prtljaga, L.~Pavesi, G.~Pucker, and M.~Ghulinyan,
  \enquote{{A fully integrated high-Q whispering-gallery wedge resonator},}
  {\protect\JournalTitle{Optics express}} \textbf{20}, 22934--22942 (2012).

\bibitem{bogaerts2012silicon}
W.~Bogaerts, P.~De~Heyn, T.~Van~Vaerenbergh, K.~De~Vos, S.~Kumar~Selvaraja,
  T.~Claes, P.~Dumon, P.~Bienstman, D.~Van~Thourhout, and R.~Baets,
  \enquote{Silicon microring resonators,} {\protect\JournalTitle{Laser \&
  Photonics Reviews}} \textbf{6}, 47--73 (2012).

\bibitem{HOM_ring}
E.~E. Hach, S.~F. Preble, A.~W. Elshaari, P.~M. Alsing, and M.~L. Fanto,
  \enquote{Scalable hong-ou-mandel manifolds in quantum-optical ring
  resonators,} {\protect\JournalTitle{Phys. Rev. A}} \textbf{89}, 043805
  (2014).

\bibitem{kaulfuss2023backscattering}
P.~L. Kaulfuss, P.~M. Alsing, R.~J. Birrittella, and D.~L. Vitullo,
  \enquote{Backscattering and hong-ou-mandel manifolds in microring
  resonators,} {\protect\JournalTitle{arXiv preprint arXiv:2305.10523}}
  (2023).

\bibitem{Azzini2012}
S.~Azzini, D.~Grassani, M.~J. Strain, M.~Sorel, L.~G. Helt, J.~E. Sipe,
  M.~Liscidini, M.~Galli, and D.~Bajoni, \enquote{Ultra-low power generation of
  twin photons in a compact silicon ring resonator,}
  {\protect\JournalTitle{Optics Express}} \textbf{20}, 23100--23107 (2012).

\bibitem{Engin2013}
E.~Engin, D.~Bonneau, C.~M. Natarajan, A.~S. Clark, M.~G. Tanner, R.~H.
  Hadfield, S.~N. Dorenbos, V.~Zwiller, K.~Ohira, and N.~e.~a. Suzuki,
  \enquote{Photon pair generation in a silicon micro-ring resonator with
  reverse bias enhancement,} {\protect\JournalTitle{Optics express}}
  \textbf{21}, 27826--27834 (2013).

\bibitem{niehusmann2004ultrahigh}
J.~Niehusmann, A.~V{\"o}rckel, P.~H. Bolivar, T.~Wahlbrink, W.~Henschel, and
  H.~Kurz, \enquote{Ultrahigh-quality-factor silicon-on-insulator microring
  resonator,} {\protect\JournalTitle{Optics letters}} \textbf{29}, 2861--2863
  (2004).

\bibitem{atvars2021analytical}
A.~Atvars, \enquote{Analytical description of resonances in fabry--perot and
  whispering gallery mode resonators,} {\protect\JournalTitle{JOSA B}}
  \textbf{38}, 3116--3129 (2021).

\end{thebibliography}

\end{document}